\documentclass[a4paper,fleqn,usenatbib]{mnras}

\usepackage{mathptmx}

\usepackage[T1]{fontenc}
\usepackage{ae,aecompl}

\usepackage{graphicx}	
\usepackage{amsmath}	
\usepackage{amssymb}	

\usepackage{natbib}

\usepackage{graphics}
\usepackage{cite}
\usepackage{float}
\usepackage{amssymb,amsmath}
\usepackage{color}
\usepackage{url}

\def\mydoubleq#1{``#1''}

\title[Evidence for Variable, Correlated X-ray and Optical/IR Extinction toward the pre-MS Binary TWA 30]{Evidence for Variable, Correlated X-ray and Optical/IR Extinction toward the Nearby, Pre-main Sequence Binary TWA 30}

\author[D. A. Principe et al.]{
David A. Principe,$^{1,}$$^{2,}$$^{3}$\thanks{E-mail: daveprincipe1@gmail.com}
G. Sacco,$^{4}$
J. H. Kastner$^{3}$
B. Stelzer$^{5}$
and J. Alcal{\'a}$^{6}$
\\
$^{1}$N{\'u}cleo de Astronom{\'i}a de la Facultad de Ingenier{\'i}a, Universidad Diego Portales, Av. Ej{\'e}rcito 441, Santiago 8320000, Chile\\
$^{2}$Millennium Nucleus Protoplanetary Disks, Chile\\
$^{3}$Laboratory for Multiwavelength Astrophysics, Rochester Institute of Technology (RIT), 1 Lomb Memorial Dr., Rochester 14623, USA\\
$^{4}$ INAF - Osservatorio Astrofisico di Arcetri, Largo E. Fermi 5,Firenze I-50125, Italy \\
$^{5}$ INAF - Osservatorio Astronomico di Palermo, Piazza del Parlamento 1, 90134 Palermo, Italy\\
$^{6}$ INAF - Osservatorio Astronomico di Capodimonte, via Moiariello 16, 80131, Napoli, Italy
}

\pubyear{2016}

\begin{document}
\label{firstpage}
\pagerange{\pageref{firstpage}--\pageref{lastpage}}
\maketitle

\begin{abstract}
We present contemporaneous XMM-Newton X-ray and ground-based optical/near-IR spectroscopic observations of the nearby ($D \approx 42$ pc), low-mass (mid-M)  binary system TWA 30A and 30B. The components of this wide (separation $\sim$3400 AU) binary are notable for their nearly edge-on disk viewing geometries, high levels of variability, and evidence for collimated stellar outflows. We obtained XMM-Newton X-ray observations of TWA 30A and 30B in 2011 June and July, accompanied (respectively) by IRTF SpeX (near-IR) and VLT XSHOOTER (visible/near-IR) spectroscopy obtained within $\sim$20 hours of the X-ray observations. TWA 30A was detected in both XMM-Newton observations at relatively faint intrinsic X-ray luminosities ($L_{X}$$\sim$$8\times10^{27}$ $erg$ $s^{-1}$) compared to stars of similar mass and age . The intrinsic (0.15-2.0 keV) X-ray luminosities measured in 2011 had decreased by a factor 20-100 relative to a 1990 (ROSAT) X-ray detection. TWA 30B was not detected, and we infer an upper limit of ($L_{X}$ $\lesssim$ 3.0 $\times$ $10^{27}$ erg s$^{-1}$).   We measured a large change in visual extinction toward TWA 30A (from $A_V \approx 14.9$ to $A_V \approx 4.7$) between the two 2011 observing epochs, and we find evidence for a corresponding significant decrease in X-ray absorbing column ($N_H$). The apparent correlated change in $A_V$ and $N_H$ is suggestive of variable obscuration of the stellar photosphere by disk material composed of both gas and dust. However, in both observations, the inferred $N_{H}$ to $A_{V}$ ratio is lower than that typical of the ISM, suggesting that the disk is either depleted of gas or is deficient in metals in the gas phase.
\end{abstract}

\begin{keywords}
stars: pre-main-sequence, circumstellar matter, low-mass, variables; T Tauri, Herbig Ae/Be, magnetic fields; planetary systems: protoplanetary disks

\end{keywords}



\section{Introduction}

Contemporaneous multiwavelength observations of nearby pre-main sequence (pre-MS) stars provide opportunities to investigate variable astrophysical phenomena related to interactions between such stars and their planet-forming circumstellar disks, as well as to measure the gas and dust content of these disks. Pre-MS stars $\leq$2 M$_{\odot}$ (i.e., T Tauri stars) are broadly subdivided into classical T Tauri stars (cTTSs) and weak-lined T Tauri stars (wTTSs) based on the strength of their H$\alpha$ emission, which serves as a proxy for their stage of evolution as star-disk systems.  During the cTTS (strong H$\alpha$ emission) stage, the pre-MS star is orbited by an optically thick circumstellar disk from which the star is actively accreting via magnetospheric accretion processes \citep{Koenigl1991,Shu1994}.  Eventually the circumstellar gas disk dissipates \citep[via photoevaporation, planet formation, and/or viscous evolution; e.g.,][]{Gorti2009,Ciesla2007,Lubow2006,Zhu2012}, the accretion process halts, and the pre-MS star enters the wTTS (weak H$\alpha$ emission) phase of evolution.  As discussed in \citet[][ and ref. therein]{Williams2011} , the median disk lifetime for pre-MS stars  (i.e., the length of the cTTS phase) is $\sim$ 3 Myr, although disk dissipation times for any individual system vary widely (i.e., $\sim$1-10 Myr).  Therefore, pre-MS stars with circumstellar disks $\gtrsim$ 8 Myr provide a unique opportunity to investigate late-stage disk evolution.

Near infrared, optical and X-ray variability has been characterized in many pre-MS star-disk systems and is the likely result of star-disk interactions \citep[e.g., magnetospheric accretion, disk warps and/or clumps, winds, hotspots;][]{Schisano2009,Flaherty2010,Bouvier2007, Principe2014}.  The variability timescale will depend on the mechanism producing such variability and can range anywhere from hours to years.  Moreover, some of these variability mechanisms are dependent on the orientation of the star-disk system with respect to our line of sight.  Previous studies of nearly edge-on or highly inclined circumstellar disks confirm the presence of variability associated with the temporary and/or periodic obscuration of the stellar photosphere from disk warps/and or clumps \citep[e.g., AA Tau, RY Lupi, T Cha;][]{Bouvier2003, Grosso2007, Manset2009, Schisano2009}. Such photospheric obscuration would only be expected to be observed in highly inclined disks, because the bulk of circumstellar material is confined to the disk plane. Contemporaneous multiwavelength observations can elucidate this mechanism of variability in star-disk systems. 

More generally, multiwavelength studies of nearby ($\lesssim$100 pc), pre-MS M stars provide the opportunity to investigate the early evolution of the most plentiful type of star in the Galaxy \citep[][and ref. therein]{Lepine2013}. High-energy radiation (e.g., NUV, FUV, X-ray) from nearby M-type stars decreases with increasing age, as \citet{Stelzer2013} demonstrated by comparing the high-energy luminosities of early M-type pre-MS stars in the TW Hydrae Association (TWA; age$\sim$8 Myr) with those of a nearby ($<$10 pc) main-sequence M dwarf sample. However, the timescale for this decline is not well determined, especially for mid- to late-M stars. Moreover, constraints on high-energy emission and variability are particularly important for theories of planet formation around M-type stars, given that their habitable zones lie at relatively small orbital radii and that such radiation can have important effects on the evaporation and chemistry of protoplanetary disks and planetary atmospheres \citep{Tarter2007, Lammer2003, Segura2005, Chadney2015, Gorti2015}.

The two components of the recently discovered wide separation binary ($\sim$3400 AU) TWA 30 \citep{Looper2010b,Looper2010a}, represent two of the nearest examples of actively accreting, M type pre-MS star-disk systems (d$\sim$42 pc).  TWA 30A and 30B are members of the $\sim$8 Myr-old TWA, one of the nearest co-moving groups of pre-MS stars to Earth \citep{Kastner1997,Mamajek2002,Ducourant2014}.  Several members of the TWA have been extensively studied \citep[e.g., TW Hya and Hen 3-600;][]{Krist2000,Kastner2002,Jayawardhana1999,Huenemoerder2007,Rodriguez2015} because their proximity and age makes them ideal objects for the study of protoplanetary systems which form in the first $\sim$ 1-10 Myrs.  Furthermore, the proximity of the TWA allows for easier characterization of very low-mass stars like TWA 30A (M5) and 30B (M4) since stars of such late spectral types would be too faint for detailed study at the $>$ 100 pc distances typical of well-studied star-forming clouds.
 
Both TWA 30A and 30B have mid-IR excesses indicative of dusty circumstellar disks \citep{Schneider2012} and both display signatures of outflows and jet emission in the form of strong forbidden emission lines (e.g., [O I], [O II], [N II], and [Si II]).  These lines display small velocity shifts with respect to rest velocity, suggesting that the outflows lie nearly perpendicular to the line of sight.  If the outflows are perpendicular to the circumstellar disk, then this orientation suggests the viewing geometry of these relatively rare late-stage,  (i.e. \mydoubleq{old} $\sim8$ Myr) disks around TWA 30A and 30B are nearly edge-on \citep{Looper2010b,Looper2010a}. The nearby TWA 30A and 30B systems hence offer the potential to study both gas and dust absorption and emission, respectively, by measuring the attenuation of coronally and/or accretion generated X-rays by circumstellar gas and  the attenuation and reprocessing of photospheric near-IR and optical emission by circumstellar dust, with little or no contamination from interstellar material.  Both TWA 30A and 30B display significant near-IR variability \citep{Looper2010b,Looper2010a}, although the mechanisms causing the variability in these stars appear to be different.  The near-IR variability of TWA 30A tends to follow the reddening vector, with typical values in the range of A$_{V}$ $\sim$ 1-9. This suggests that a changing column density of circumstellar disk dust is responsible for variable reddening of the stellar photosphere.  In contrast, the near-IR variability of 30B tends to follow the CTTS locus, possibly indicative of reprocessed starlight emitted by an inner-disk structure and variably scattered towards us.

TWA 30A was reported in \citet{Looper2010b} to have been detected in X-rays in the ROSAT All Sky Survey (RASS) Faint Source Catalog at a distance of 17.6 arcseconds from the 2MASS position, well within the $\sim$25 arcsecond spatial resolution of ROSAT at $\sim$1 keV.  TWA 30B was not detected in RASS.  X-rays in pre-MS stars can be coronally generated and/or the result of star-disk interactions (e.g., star-disk magnetic reconnection events and magnetically funneled accretion) and have been shown to be variable on timescales of hours to years \citep{Bouvier2007, Principe2014}.  Therefore X-ray observations of such stars and their X-ray variability offer a means to probe pre-MS stellar magnetic and accretion activity. Furthermore, \mydoubleq{soft} ($<$ 1 keV) X-ray emission is absorbed by circumstellar gas \citep[mainly atomic C, N, and O;][]{Wilms2000} and can thus be used to estimate a line of sight gas column density due to the disk.  X-ray (and UV) emission is also an important source of disk ionization and chemistry in protoplanetary disks \citep{Glassgold2012}.  X-ray (and UV) heating can alter the disk structure by heating the surface layers of the disk resulting in a \mydoubleq{flared disk} \citep{Meijerink2012}.  

To investigate the nature, origins, and variability of the X-ray emission from TWA 30A and 30B, we have obtained near-simultaneous X-ray and near-IR spectroscopy of these two systems. We present the observations in Section 2. We discuss results in Sections 3 and 4 and present our summary and conclusions in Section 5.

\section{Observations}

\subsection{X-rays}
X-ray observations of TWA 30A and 30B (Table \ref{twa30_obs}) were obtained with the European Photon Imaging Camera (EPIC) instrument on board the XMM-Newton Observatory (Obs.IDs 0672500201 and 0672500301).  Two $\sim25$ ks exposures were obtained near simultaneously with the VLT and IRTF observations, respectively.  All observations were reduced using the ESA XMM Science Analysis System (SAS) version 11.0.  Observations with the EPIC instrument include both the pn and the two MOS cameras. Observations using all three detectors have been energy filtered (0.15 - 10 keV) and time filtered for flaring particle background.  The XMM-Newton observation obtained in 2011 June was contaminated by a large solar flaring event $\sim$15 ks into the observation.  In order to extract data from this observation, the EPIC-pn event files were time filtered such that only the data during the first $\sim$15 ks were used. EPIC-MOS1 and MOS2 data from the 2011 June epoch were not used due to poor quality from the flaring particle background.  A small background flaring event ($\sim$5 ks) was time-filtered from the 2011 July observation.  The resulting effective useful exposure times for the EPIC observations are listed in Table 1. Standard XMM-SAS tasks were used to extract spectral files as well as region statistics, including source count rates and background count rate estimates at the positions of TWA 30A and 30B in both observations.  Source extraction regions were circles with radii of $\sim$15$''$ centered at the 2MASS position of TWA 30A and 30B and background extraction regions were circles with radii of $\sim$50$"$ positioned nearby TWA 30A and 30B on the same chip with no clear point sources or extended emission contaminants.

\subsection{Near Infrared and Optical}

Near-infrared/optical spectroscopy of TWA 30A and 30B was obtained in 2011 June and 2011 July with the 3.0m Infrared Telescope Facility (IRTF) and the 8.2m Very Large Telescope (VLT), respectively.  These observations were intended to obtain near-IR spectroscopic data contemporaneous with the XMM observations described in section 2.1. Near-IR spectroscopy of TWA 30A and 30B were obtained using the medium resolution Spectrograph and Imager (SpeX) on the IRTF.  Observations of TWA 30A were obtained in cross-dispersed mode (SXD) with a 0.5$''$ slit resulting in a resolution of $\sim$1200.  TWA 30B observations were obtained in prism mode with a 0.5$''$ slit resulting in a resolution of $\sim$120. The IRTF SpeX data were reduced and combined using Spextool\footnote{\url{http://irtfweb.ifa.hawaii.edu/~spex/Spextool_v3.4.txt} and a custom IDL code by D. Looper following methods described in detail in \citet{Looper2010b,Looper2010a}}.  Near-IR, optical and UV spectra of TWA 30A and 30B were obtained using the X-SHOOTER instrument on the VLT (Program 287.C-5039) in nodding mode with slit widths of 0.8, 0.9 and 0.9$''$ with corresponding resolutions of $\sim$ 6200, 8800, and 5300 for the UVB, VIS, and NIR arms, respectively.  Basic data reduction was performed using the X-SHOOTER pipeline and the subsequent telluric correction and NIR response function were performed with IRAF. More details of the XSHOOTER data reduction can be found in \citet{Alcala2014}. Details of the observations presented here are provided in Table \ref{twa30_obs}.

\begin{table*}
\centering

\caption{TWA 30 Observations \vspace{2.5mm}}

\begin{tabular}{| c | c | c | c | }
\hline

	Object & Obs. Date [UTC] & Telescope & Exposure (s)  \  \\ \hline
	TWA 30A & 2011-06-07 10:40:52 & XMM-EPIC  & 13950\footnotemark[1]/--/-- \\ 
	TWA 30A & 2011-06-08 05:53:04 & IRTF-SpeX  & 1440 \\ 
	TWA 30A & 2011-07-15 00:00:07 & VLT-XSHOOTER  & 920/560/600\footnotemark[2]  \\ 
	TWA 30A & 2011-07-15 21:59:40 & XMM-EPIC  & 14720\footnotemark[1]/21290/21320 \\ \hline
	
	TWA 30B & 2011-06-07 10:40:52 & XMM-EPIC  & 13950\footnotemark[1]/--/-- \\ 
	TWA 30B & 2011-06-08 06:35:21 & IRTF-SpeX  & 750 \\ 
	TWA 30B & 2011-07-15 00:38:27 & VLT-XSHOOTER  & 1960/1600/1160\footnotemark[2]  \\ 
	TWA 30B & 2011-07-15 21:59:40 & XMM-EPIC  & 14720\footnotemark[1]/21290/21320 \\ \hline
	
\end{tabular}

\vspace{0.5mm}

\footnotemark[1]{ Effective exposure times (pn/MOS1/MOS2) due to interruption by background flaring events. }
\footnotemark[2]{ Exposure time for each spectrograph arm in the format of UVB, VIS, NIR.  }
\label{twa30_obs}
\end{table*}

\section{Results}

\subsection{Near-IR and Optical Spectroscopy}

\begin{figure}
\centering

\includegraphics[scale=0.45]{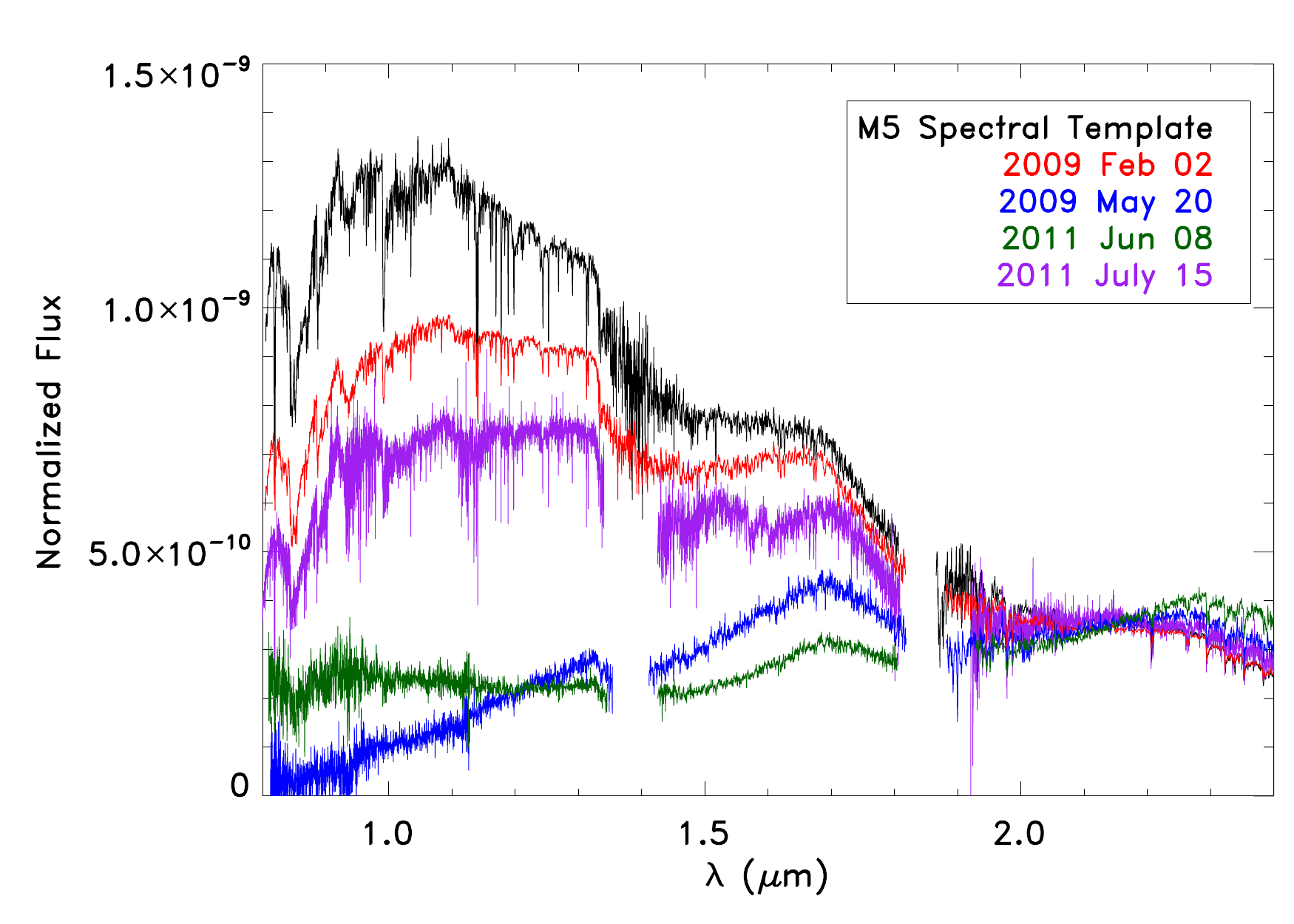}

\caption{\label{twa30A_compare}  Two extreme examples of the TWA 30A near-IR SpeX spectra originally presented in \citet[][]{Looper2010a} compared to 2011 June and 2011 July SpeX (green) and VLT (purple) observations presented in this work.  An unreddened M5 spectral template is shown in black.  All spectra have been normalized to the flux at $\sim$2.15 $\mu$$m$. Spectra from top to bottom at 1 $\mu$$m$ are: M5 spectral template; 2009 Feb 02; 2011 July 15; 2011 Jun 08; 2009 May 20. }  
\end{figure}

\begin{figure}
\centering
\includegraphics[scale=0.45]{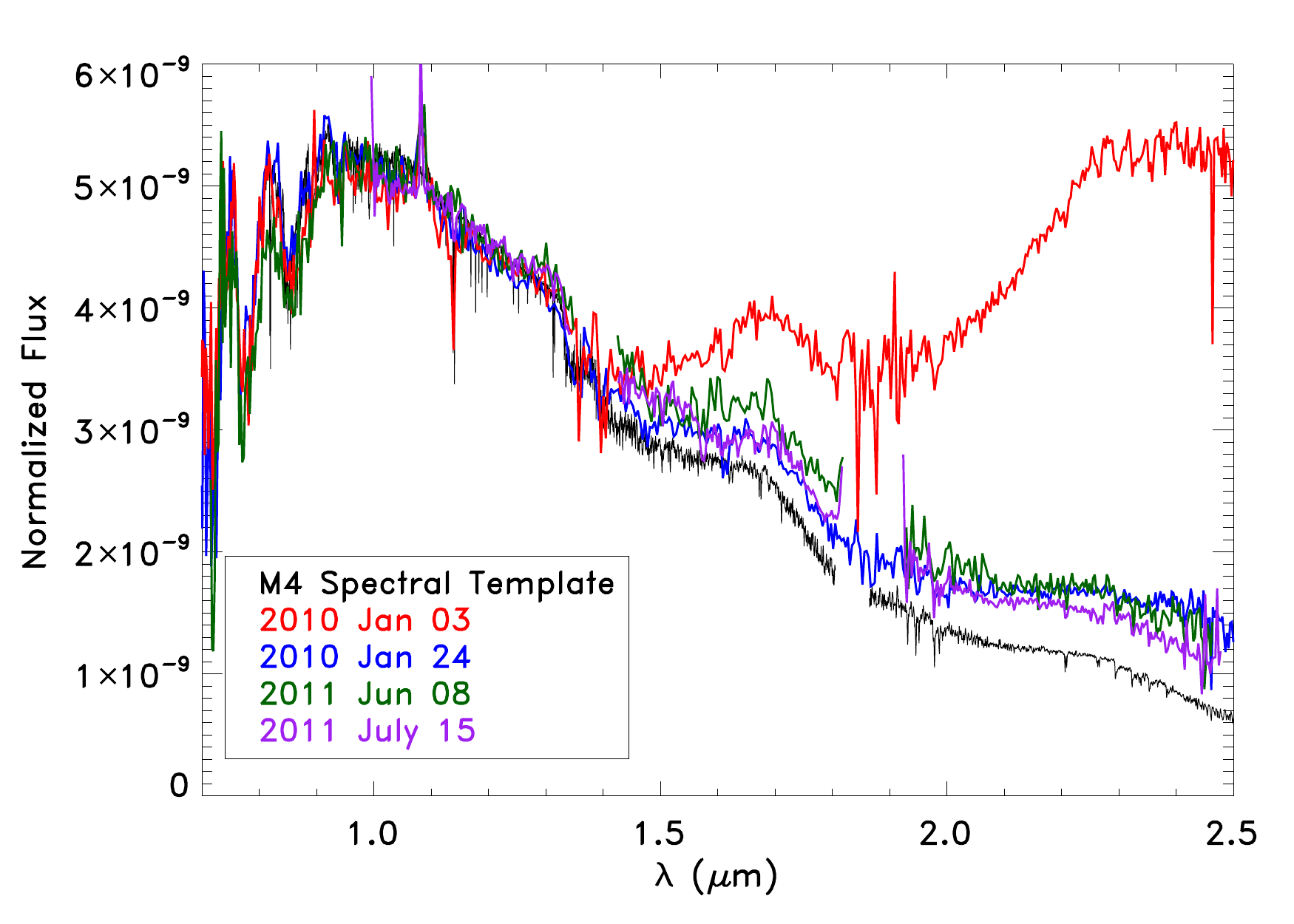}

\caption{\label{twa30B_compare} Two extreme examples of TWA 30B SpeX observations originally presented in \citet[][]{Looper2010b} exhibiting high and low levels of disk excess compared to an unreddened M4 spectral template (black) and 2011 June and 2011 July SpeX (green) and VLT (purple) observations presented in this work.  The VLT spectrum of TWA 30B has been rebinned to match the resolution of the IRTF SpeX spectra. All spectra have been normalized to the flux at $\sim$1.27 $\mu$$m$.  The spectra at epochs 2010 Jan 24, 2011 Jun 08 and 2011 July 15 are nearly coincident.}  
\end{figure}

Reduced spectra of TWA 30A and 30B from \citet{Looper2010b,Looper2010a} were provided by D. Looper and compared to the more recent IRTF and VLT observations presented in this work (Section 2.2; Figures \ref{twa30A_compare} and \ref{twa30B_compare}).  In the case of TWA 30A, the continuum shapes $\gtrsim$ 1.5 $\mu$$m$ during the 2011 June and July epochs are similar to those of the most and least reddened spectra presented in \citet{Looper2010b}, respectively (Figure \ref{twa30A_compare}).  We note that the TWA 30A spectra from the 2011 June and July epochs are flatter in the J band region than the spectra displayed in \citet{Looper2010b}. In contrast to the 2011 June/July continuum shape change in TWA 30A, both TWA 30B spectra appear to closely resemble those in \citet{Looper2010a} that were obtained during periods where this object displayed minimal near-IR excess (Figure \ref{twa30B_compare}). The comparison between previous near-IR spectroscopy and the near-IR spectra we obtained is discussed further in Section 4.

\begin{figure}
\centering
\includegraphics[scale=0.45]{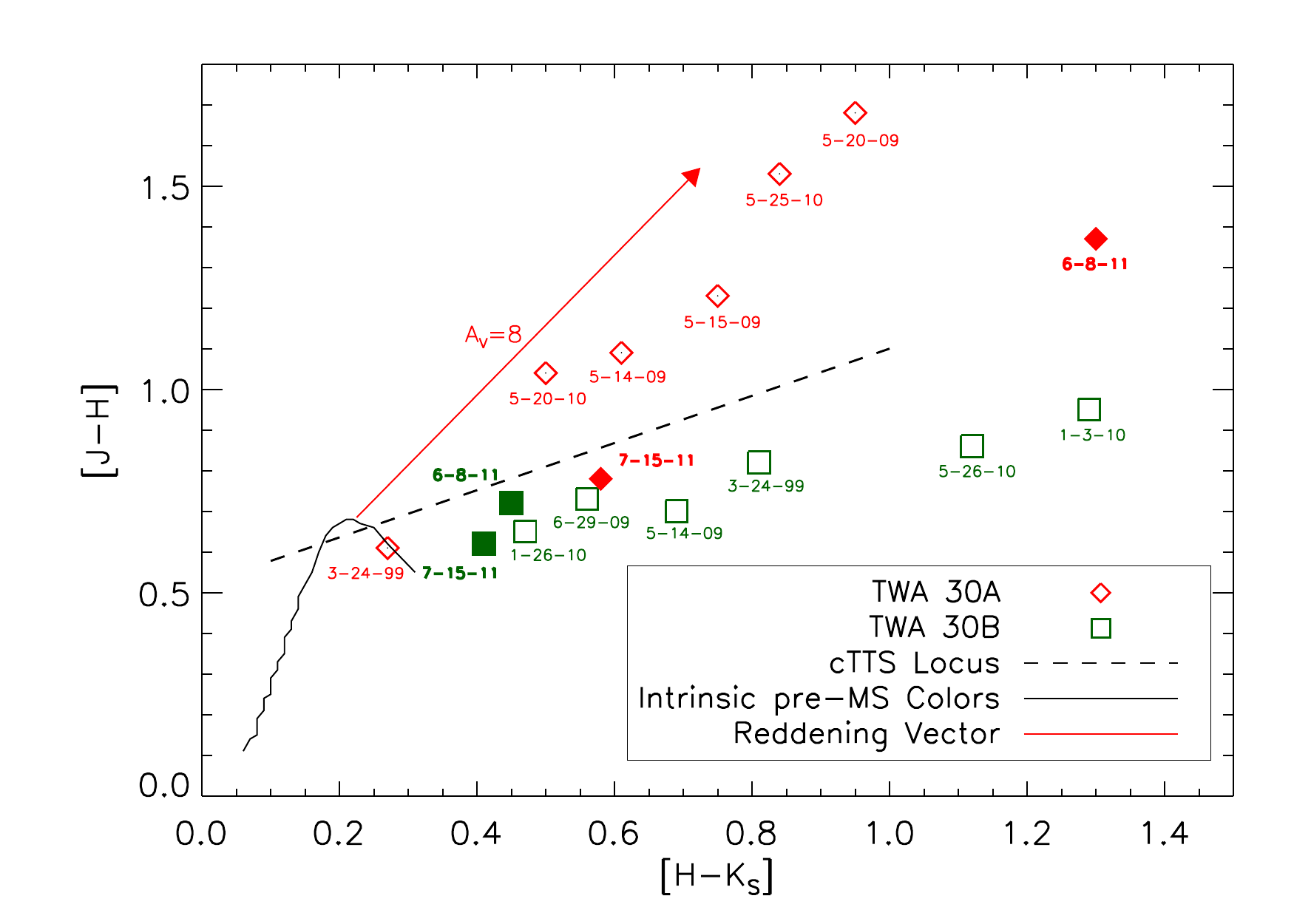}
\caption{\label{2mass_cc} JHK color-color plot of the variable TWA 30A (red diamonds) and 30B (green squares). A selection of photometry extracted from spectral observations between 1999-2010 and initially presented in \citet{Looper2010b} are shown with unfilled symbols.   Photometry extracted from IRTF SpeX and VLT spectral observations in 2011 are shown with filled symbols.  The intrinsic pre-MS colors of $\sim$5-30 Myr old stars are shown in black \citep{Pecaut2013}.  The classical T Tauri star locus \citep{Meyer1997} and a reddening vector corresponding to an A$_{V}$=8 are shown in black (dashed) and red (solid) lines, respectively.  }  
\end{figure}

\begin{figure}
\centering
\includegraphics[scale=0.45]{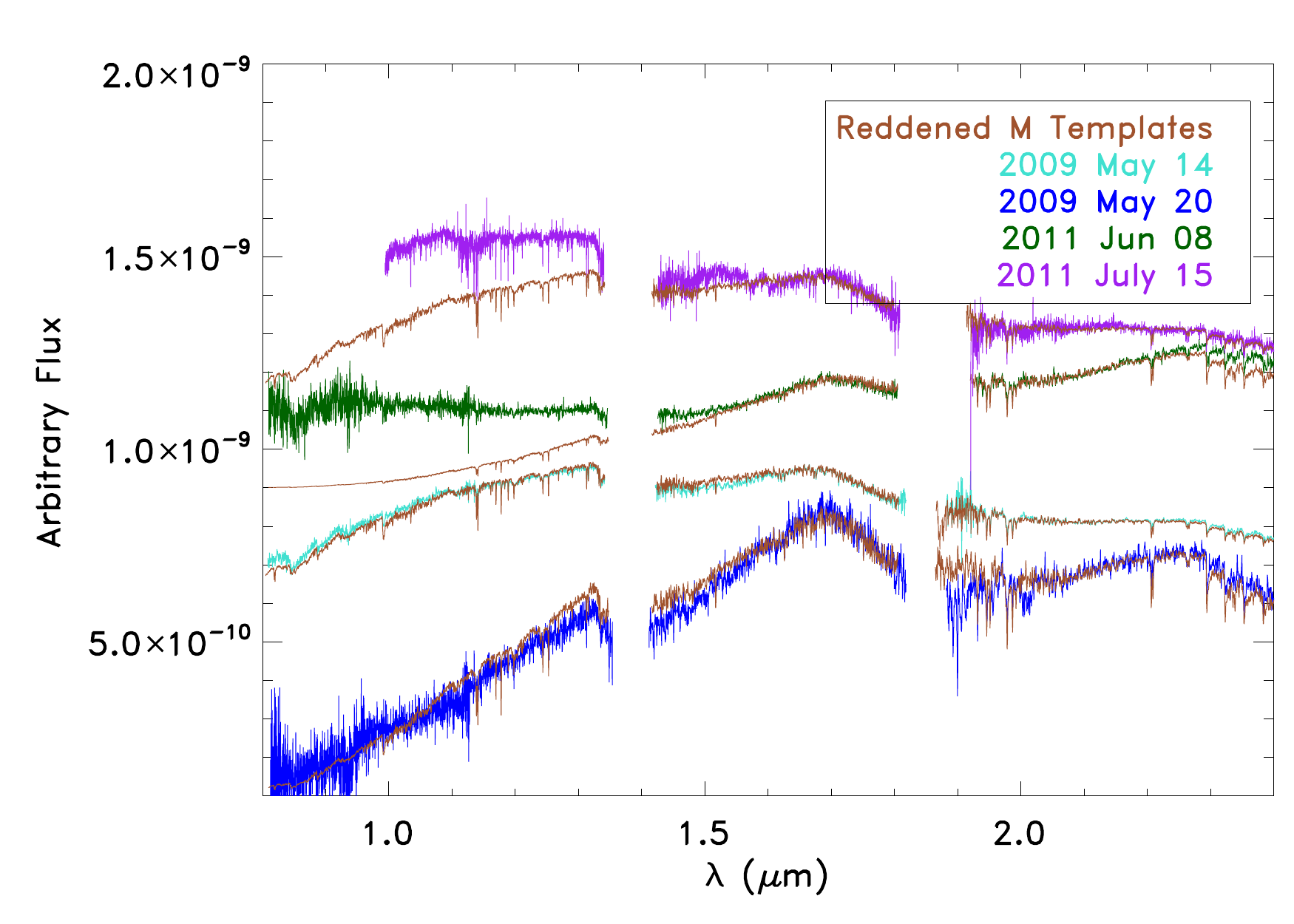}
\caption{\label{twa30A_red}  A comparison between a selection of extincted SpeX observations of TWA 30A presented in \citet[][blue and teal]{Looper2010b} and this work (green and purple). All spectra were fit with reddened M templates ($R_{V}=3.1$; M5 for epochs  2009, May 14 and 2011, July 11 and M6 for 2009, May 20 and 2011, June 8) with corresponding reddening values of $A_{V}$$=4.65$, $A_{V}$$=4.7$, $A_{V}$$=9.0$ and $A_{V}$$=14.9$ , respectively. Each spectrum was normalized to its corresponding spectral template at $\sim$2.15 $\mu$$m$. An excess is clearly visible in the 2011 June and July epochs which was not previously observed. Spectra from top to bottom at 1.5 $\mu$$m$ (with coincident reddened templates) are: 2011 July 15, 2011 Jun 8, 2009 May 14, 2009 May 20.}

\end{figure}

Standard broad-band near-IR (JHK) photometry were extracted from the 2011 June and July IRTF and VLT spectra using standard 2MASS JHK filter bandpasses \citep{Cohen2003}.  A custom IDL program was used to integrate spectral flux within each filter bandpass\footnote{The near-IR spectroscopy results presented in this work will be limited to calculations involving extinction and continuum shape.  Detailed UVB-NIR line analyses will be presented in a future paper by B. Stelzer.}. In Figure \ref{2mass_cc} the 2011 June/July effective 2MASS colors are shown in a near-IR color-color diagram and compared to a selection of data presented in \citet{Looper2010a}. In the case of TWA 30A, Figures \ref{2mass_cc} and \ref{twa30A_red} demonstrates a large variability in extinction and a previously unobserved near-IR excess ($\sim$0.8-1.6 $\mu$$m$).  Due to the near-IR excess apparent in the data presented here, we choose not to derive $L_{bol}$ from J band photometry and instead adopt the previously derived $log(L_{bol}/L_{\odot})=-1.70$ from \citet{Looper2010b} with the assumption that $L_{bol}$ has not changed considerably between the two epochs.  TWA 30B displays levels of near-IR excess near or below the minimum previously reported.

Optical extinction ($A_{V}$) was measured for TWA 30A following the same procedure described in \citet{Looper2010b}: i.e., spectral templates were reddened according to the reddening curve presented in \citet{Cardelli1989} with $R_{V}$=3.1, and the resulting reddened template spectra were then compared by eye to the observed spectra so as to estimate the best match in terms of both $A_{V}$ and spectral type.  We spectroscopically classified TWA 30A for each observation with the following spectral templates originally used in \citet{Looper2010b} from the IRTF Spectral Library \citep{Rayner2009}: Gl 213 (M4 V), Gl 51 (M5 V) and Gl 406 (M6 V).  The best match to the 2011 June observation of TWA 30A was provided by the M6 spectral template with an extinction correction of $A_{V}$ $=$ 14.9. The best match to the 2011 July observation was provided by the M5 spectral template with an extinction correction of $A_{V}$ $=$ 4.7 (Figure \ref{twa30A_red}). 

Extinction and errors were estimated for TWA 30A by adjusting the reddening ($A_{V}$) of the template to obtain the best visual match to the spectrum in the 1.6 - 2.0 $\mu$$m$ wavelength region. The observed spectra did not match the spectral templates at wavelengths $\lesssim$ 1.6$\mu$$m$, with the comparison indicating the presence of a continuum excess in this wavelength range (Figure \ref{twa30A_red}).  Therefore, measurements of $A_{V}$ presented here assume the 1.6 - 2.0 $\mu$m wavelength region is relatively excess-free.  This assumption is supported by evidence in that this part of the spectrum fits the reddened spectral templates well in both the 2011 June and July  observations and none of the previous eight near-IR observations of TWA 30A presented in \citep{Looper2010b} display an excess at this wavelength.  In order to consider dust size distributions different from that of the ISM, we also determined extinction assuming (ISM-like and non-ISM-like) values of $R_{V}$=2.75 and 5.3.  For these values of $R_{V}$ we find, respectively, estimated optical extinction ($A_{V}$) values of 17.3 and 12.7 for the June epoch, and values of 5.1 and 3.7 for the July epoch.

The best match to the TWA 30B spectra was provided by the M4 spectral template (Figure \ref{twa30B_compare}). The spectral types for TWA 30A and 30B determined in this work are consistent with those previously reported in \citet{Looper2010b,Looper2010a}. No values of $A_{V}$ were calculated for TWA 30B because variability in this system is likely due to near-IR excess and not photospheric extinction \citep{Looper2010a}.  The JHK colors and measured values of $A_{V}$ for TWA 30A and JHK colors for TWA 30B are listed for the 2011 June and July epochs in Table \ref{twa30_jhk}. The optical extinction measured from the 2011 June epoch ($A_{V}$ $=$ 4.7) for TWA 30A is consistent with the 2009 May 14-15 measurements reported in \citet{Looper2010b}. Our measured extinction for the 2011 June epoch ($A_{V}$ $=$ 14.9) is significantly higher than the highest extinction previously reported \citep[$A_{V}$ $=$ 9.0 in 2009 May 20;][]{Looper2010b}.

\begin{table}
\centering

\caption{J H K Colors and Extinction \vspace{2.5mm}}

\begin{tabular}{| c  c  c  c  c  c |  }

\hline
Object  & Obs. Date & [J-H] & [H-K] & [J-K] &  A$_{V}$\footnotemark[1] \\ \hline
	TWA 30A & 2011-06-08 & 1.37 & 1.3 & 2.67 & 14.9$^{+0.6}_{-1.3}$ \\ 
	TWA 30A & 2011-07-15 & 0.78 & 0.58 & 1.36 & 4.7$^{+0.6}_{-0.6}$ \\ 
	TWA 30B & 2011-06-08 & 0.72 & 0.45 & 1.17 & - \\ 
	TWA 30B & 2011-07-15 & 0.62 & 0.41 & 1.03 &  - \\ \hline
		
\end{tabular}
\vspace{0.5mm}

\footnotemark[1]{Measured using reddened template spectra with $R_{V}=3.1$}.

\label{twa30_jhk}
\end{table}

\subsection{X-ray Properties of TWA 30A and 30B}

\begin{figure}
\centering
\includegraphics[scale=0.45]{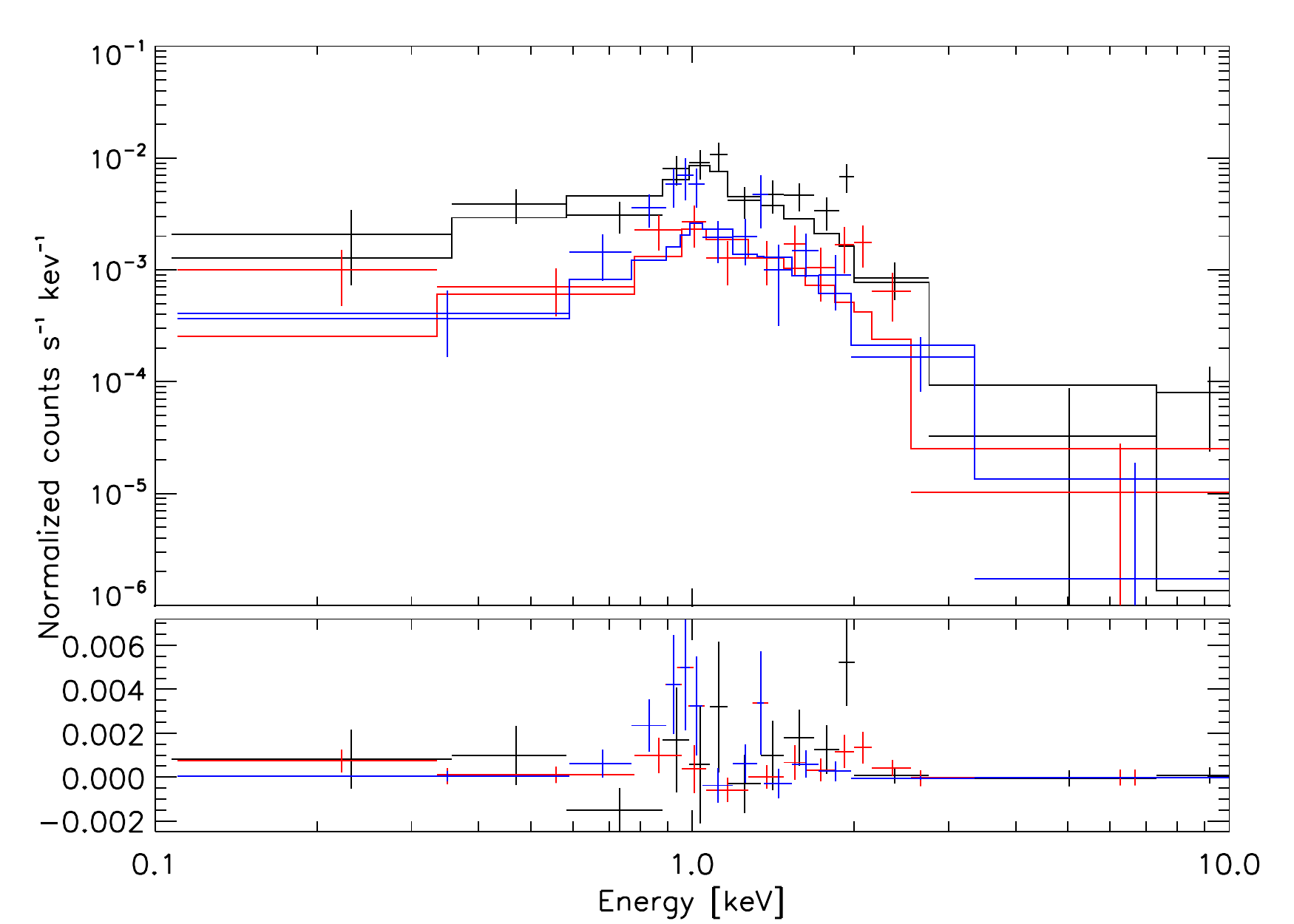}
\caption{\label{twa30a_xmmfit} The 2011 July X-ray spectrum of TWA 30A (crosses) fit with an absorbed one-temperature thermal plasma model (histogram).  Each color (black, blue, red) represents the same model fit to data from each of the XMM-Newton EPIC detectors (pn, MOS1, MOS2), respectively. }
\end{figure}

X-ray spectra of TWA 30A from the two XMM-Newton observations were extracted and fit with an absorbed, single-temperature thermal plasma emission model (apec) assuming a solar abundance of 0.2 using XSPEC\footnotemark.  The metallicity of 0.2 solar was adopted for consistency with previous spectral fitting of XMM data for the M2.5 pre-MS star TWA 11B \citep{Lopez2007}.  The free parameters of the fit were model normalization (proportional to the emission measure), X-ray plasma temperature ($T_{X}$), and neutral hydrogen absorbing column ($N_{H}$). The absorbed X-ray flux ($F_{X}$) was determined from the best fit model (red. $\chi^{2}$ = $1.42$ and $0.92$ for the 2011 June and July epochs, respectively). An intrinsic (i.e., unabsorbed) X-ray luminosity ($L_{X}$) was calculated using the best fit model assuming $N_{H}$=0 and a distance of 42 pc \citep{Looper2010a}.  The resulting X-ray spectral fit parameters for TWA 30A between 0.15 - 10.0 keV are listed in Table \ref{xray_par}. We find no evidence of excess soft ($\lesssim$ 1 keV) X-ray emission indicative of shocked-heated plasma --such as would be expected given the presence of accretion shocks or shocks in the acceleration region of a jet-- at either observing epoch.  The 2011 July X-ray spectrum of TWA 30A and best-fit model is shown in Figure \ref{twa30a_xmmfit}.  An attempt to better constrain the X-ray flux near 1 keV was made assuming Ne and Fe abundances similar to those determined in the case of TW Hya \citep[Ne=2.5, Fe=0.2;][]{Brickhouse2010}. However, such a model does not improve the fit and yields similar values for the X-ray temperature, absorbing column, and luminosity.  Hence, we report the results of the fit to the 0.2 solar abundance apec model (Table \ref{xspec}).

\begin{figure}
\centering
\includegraphics[scale=0.43]{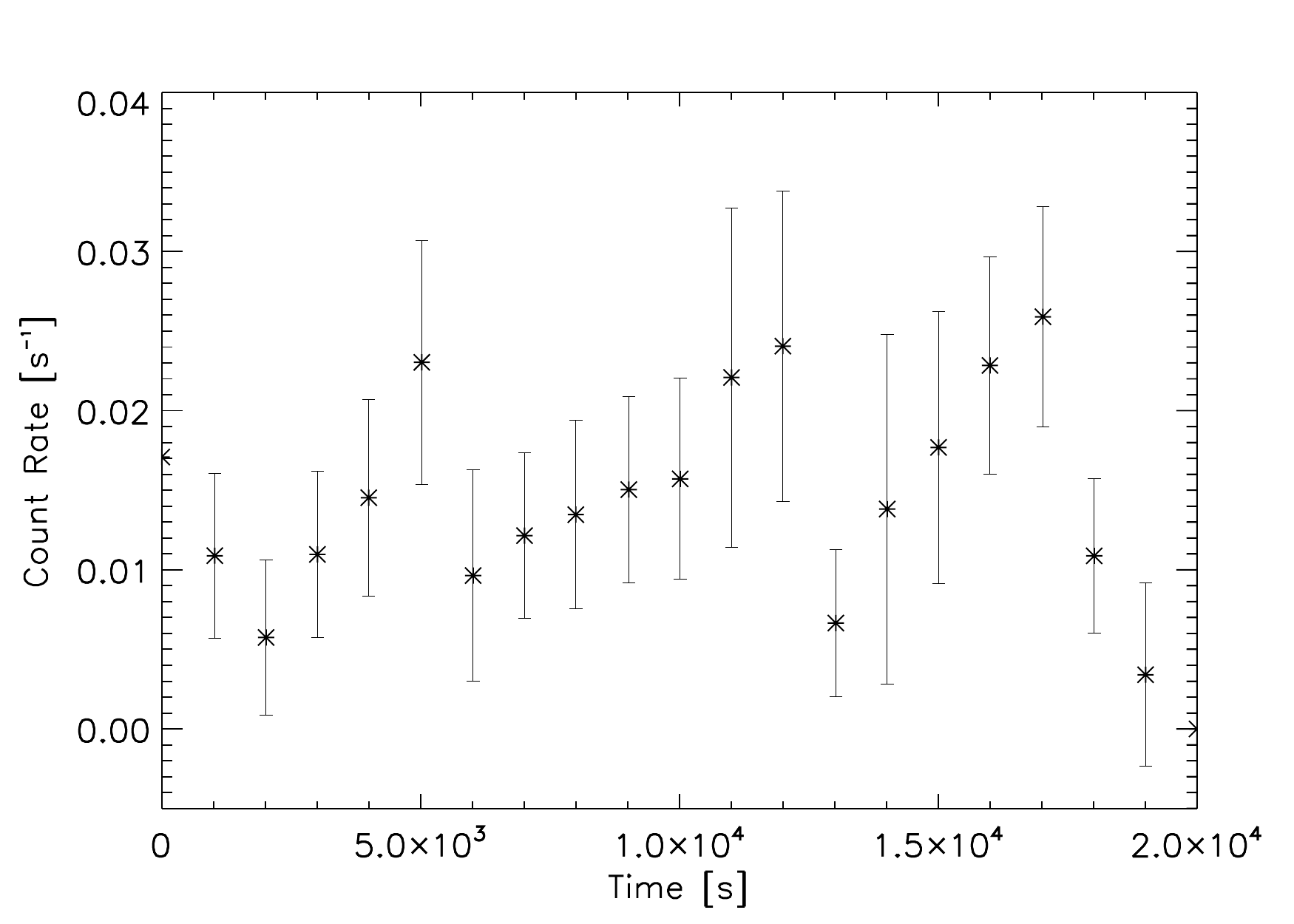}
\caption{\label{twa30_xmm_lc} The 2011 July XMM-Newton X-ray light curve of TWA 30A with one sigma error bars. }
\end{figure}

The 2011 June TWA 30A X-ray EPIC-pn spectral fit suffered from large and unconstrained errors in the fit parameters (e.g., $kT$ and $F_{X}$) due to a low count rate and particle flaring contamination, which reduced the effective exposure time.  The 2011 July X-ray spectral fit is better constrained. X-ray light curves were also constructed for TWA 30A using the 2011 July data, to search for short-timescale X-ray variability (Figure \ref{twa30_xmm_lc}).  Low levels of flaring can be seen in TWA 30A (e.g., increase in count rate by factor of $\sim$2).  X-ray variability between the 2011 June and 2011 July epochs is not discussed due to the poorly constrained plasma temperature in the 2011 June epoch.

 As reported in the RASS Faint Source Catalog and \citet{Looper2010b}, TWA 30A was detected with ROSAT in 1990 with a count rate of (2.5 $\pm$ 1.0) $\times 10^{-2}$ counts s$^{-1}$.   We estimate the 0.1-2.0 keV intrinsic X-ray luminosity associated with the ROSAT count rate using \footnotetext{http://heasarc.gsfc.nasa.gov/xanadu/xspec/} HEASARC PIMMS\footnote{https://heasarc.gsfc.nasa.gov/cgi-bin/Tools/w3pimms/w3pimms.pl} assuming the same spectral parameters (e.g., $kT$ and $N_{H}$) determined from the XMM-Newton best-fit models.  We estimate two values of intrinsic $L_{X}$ for the 1990 RASS observation corresponding to the high-extinction and low-extinction  states observed in 2011 June and July, respectively (Table \ref{xspec}). These parameters result in a ROSAT intrinsic (i.e., unabsorbed) X-ray luminosity of $5.2 \times 10^{29}$ erg s$^{-1}$ and  $1.4 \times 10^{29}$ erg s$^{-1}$, respectively.   These values of $L_{X}$ differ significantly from those derived from the same count rate and reported in \citet[][ $L_{X}=4.0\times10^{28}$ erg s$^{-1}$]{Looper2010b} because they report an X-ray luminosity estimated from the ROSAT count rate with a method derived for main sequence low-mass dwarf stars within 7pc of the Earth \citep[i.e., sources with negligible ISM and no circumstellar absorption;][] {Schmitt1995}.  Therefore, their calculation does not consider the effects of absorption from circumstellar material which, in the case of a nearly edge-on disk like TWA 30A, can be significant (Table \ref{xray_par}). Thus, we adopt our estimated values of ROSAT $L_{X}$. In order to compare the 2011 June/July XMM observations with that of ROSAT, we recalculate the intrinsic X-ray luminosities in the 0.15-2.0 keV energy bandpass.  We derive 0.15-2.0 keV X-ray luminosities of $L_{X} = 5.3\times10^{27}$ and $L_{X} = 6.2\times10^{27}$ erg s$^{-1}$ for the 2011 June and July XMM data, respectively.  We conclude that the 0.15-2.0 keV X-ray luminosity of TWA 30A decreased by a factor of $\sim$20-100, where the precise factor decrease depends on the value of $N_{H}$ assumed for the 1990 (ROSAT) data.

TWA 30B was not detected with XMM-Newton in either observation; the upper-limit on its pn count rate is estimated to be $\lesssim$ 1.6 x 10$^{-3}$ s$^{-1}$ for the 2011 July epoch.  Assuming D=44 pc and the same X-ray spectral fit parameters as determined for TWA 30A in its high-extinction state in 2011 June (Table \ref{xray_par}), we determine an upper-limit on the X-ray luminosity of TWA 30B of $L_{X}$ $\lesssim$ 3.0 $\times$ $10^{27}$ erg s$^{-1}$.

\begin{table*}
\footnotesize
\centering
\caption{X-ray Spectral Fit Parameters for TWA 30A \label{xspec} \vspace{2.mm}}

\begin{tabular}{| c | c | c | c | c | c | c | c |}
\hline
	Obs. Date & Count Rate & pn Exp. Time & $N_{H}$ & kT & Norm. & Absorbed $F_{X}$ & Intrinsic $L_{X}$\footnotemark[1]    \\ 
	 UTC &   [ks$^{-1}$] & [ks$^{-1}$] & [$1 \times 10^{22} $ cm$^{-2}$] & [keV] & $\times 10^{-5}$   & [erg cm$^{-2}$ s$^{-1}$] & 0.15-10.0 keV [erg s$^{-1}$] \\ \hline
	2011-06-08 & 5.1 $\pm{1.8}$\footnotemark[2] & 14.0 & 0.95$_{-0.61}^{+1.75}$ & 2.22\footnotemark[3] & 4.19$_{-1.70}^{+2.76}$ & \footnotemark[3]5.59E-14$_{}^{+1.80E-14}$ & 1.7E+28 \vspace{1.5mm} \\ 
       2011-07-15 & 11 $\pm{1.2}$ & 14.7 & 0.16$_{-0.06}^{+0.10}$ & 1.6$_{-0.25}^{+0.30}$ & 3.43$_{-0.66}^{+0.83}$ & 2.20E-14$_{-2.9E-15}^{+3.6E-15}$ & 7.9E+27	\\ \hline
	
\end{tabular}

\vspace{0.5mm}

\footnotemark[1]{See section 3.2 for 0.15-2.0 keV Intrinsic $L_{X}$ during the 2011 June/July epochs for comparison with the ROSAT 1990 detection.}
\footnotemark[2]{X-ray spectral fitting for 2011 June was performed using only the EPIC-pn detector.}
\footnotemark[3]{Poorly constrained parameter due to low number of counts.}

\label{xray_par}
\end{table*}

\begin{figure}
\centering
\includegraphics[scale=0.45]{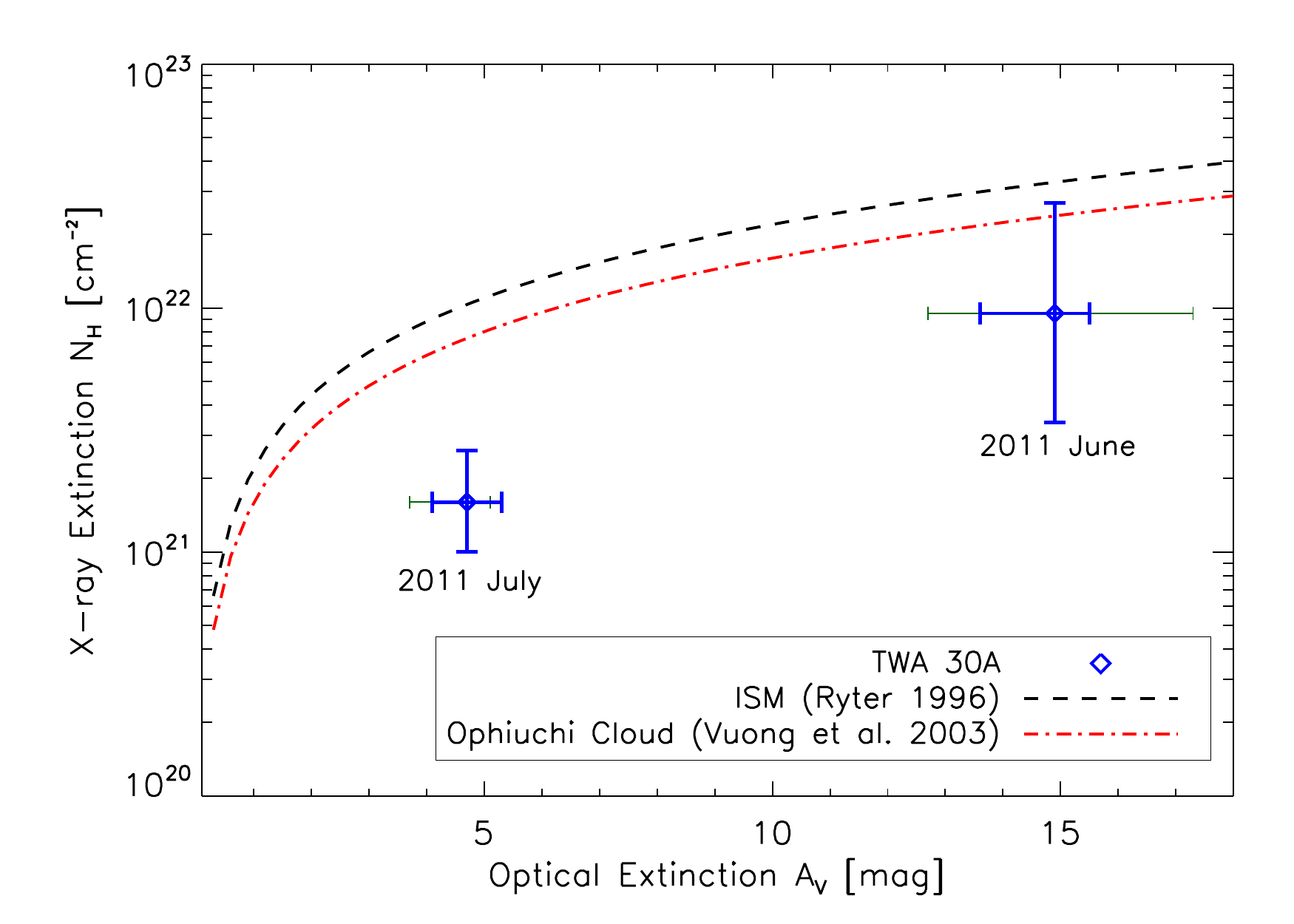}
\caption{\label{twa30_nh_av} The relationship between gas ($N_{H}$) and dust ($A_{V}$) extinction in TWA 30A for two epochs of contemporaneous observations compared to the ISM \citep[black dashed line;][]{Ryter1996} and the $\rho$ Ophiuchi molecular cloud \citep[red dash-dotted line;][]{Vuong2003}, all of which assume an $R_{V}=3.1$. For each observation, the uncertainty in the determinations of $A_{V}$ corresponding to a plausible range of $R_{V}$ for dust in the TWA 30A disk (i.e., $R_{V}$ = 2.75 to 5.3; see Sections 3.1, 4.2) is shown as a green error bar. }
\end{figure}

\section{Discussion}

\subsection{Near Infrared and X-ray Variability}

\citet{Looper2010b,Looper2010a} had previously shown, based on near-IR spectroscopy collected over the period 2008-2009, that TWA 30A and 30B are highly near-IR variable on timescales of days and that their variations follow the reddening vector and the cTTS (dust emission excess) locus, respectively.  For TWA 30A, this variable emission was suggested to be due to variable obscuration of the stellar photosphere by circumstellar dust.  This explanation can account for the variable reddening detected.  However, in the observations presented in this work, reddening of the photosphere alone cannot account for TWA 30A's variability.  Instead, our measurements of [J-H] and [H-K] colors (Figure \ref{2mass_cc}) evidently display a combination of both reddening and $\sim$0.8-1.6 $\mu$m near-infrared excess.  Furthermore, the slope of the continuum between $\sim$0.8-1.6 $\mu$$m$ is flatter in the 2011 June and July observations compared with those of the \citet{Looper2010a} observations, possibly indicative of an additional dust scattering component or a continuum excess (Figure \ref{twa30A_red}). In the case of classical T Tauri stars, near-IR continuum excess is generally indicative of accretion \citet{Hartigan2003}. In the case of TWA 30A, however, accretion is unlikely to contribute excess flux in the 1.6 - 2.0 $\mu$m region where our reddened spectral template fits assume an excess-free region (Section 3.1). We caution that strong accretion signatures, including veiling affecting the Li I and K I lines, have been previously observed in TWA 30A \citep{Looper2010b} and in some cases, cTTSs have demonstrated NIR continuum excess in the entire $\sim$0.48-2.1 $\mu$m range \citep{Fischer2011}.  Nonetheless, the presence of a slight 2.2-2.4 $\mu$m excess in the 2011 June observation indicates that excess near-IR continuum emission alone cannot account for the shape of this spectrum.  Both the 2011 June and July near-IR spectral observations of TWA 30B closely match the colors and spectral shape of the previous \citep{Looper2010a} observations obtained at the lowest levels of near-IR excess (Figures \ref{twa30B_compare} and \ref{2mass_cc}).

In the case of TWA 30A, a potential decrease of 1-2 orders of magnitude in $L_{X}$ over two decades is evident when comparing the results of the 1990 ROSAT and 2011 XMM observations (Section 3.2). The physical mechanism by which a pre-MS star can decrease in X-ray luminosity by such significant levels within a timespan of $\sim$20 years is unclear.  One possibility is that TWA 30A was undergoing a flaring event in 1990.  While individual flaring events for pre-MS stars may happen on timescales of hours to days \citep[e.g.,][]{Getman2005,Favata2005,Stelzer2007b}, longer-term (year-timescale) X-ray variability in count rate has been observed for pre-MS stars in L1630 \citep[e.g.,][]{Principe2014}.  Another possibility is that TWA 30A may have undergone an accretion outburst similar to those seen in FU Ori and EX Lupi type objects \citep[for a review, see][]{Audard2014}.  Such outbursts can be accompanied by order of magnitude changes in brightness in the near-IR, optical and X-ray, as in the case of V1647 Ori \citep{Kastner2006}.  However, this scenario cannot be confirmed in the case of TWA 30A as there are no optical/IR observations available contemporaneous with the 1990 ROSAT observation.  While the mechanism by which TWA 30A has decreased in $L_{X}$ remains unclear, the impact of such a change in $L_{X}$ may be significant when considering processes such as disk ionization \citep{Gorti2009,Cleeves2015} and mass-loss via X-ray induced photodissociation \citep{Owen2012} over the lifetime of the disk. 

In the context of other pre-MS stars of similar mass ($\sim$0.1 $M_{\odot}$), the intrinsic X-ray luminosity of TWA 30A as measured by ROSAT in 1990 appears typical when compared to, e.g., low mass stars in the ONC \citep[][]{Preibisch2005a} and IC 348 \citep[][]{Stelzer2012}.  However, it then follows that the significant decrease in $L_{X}$ from 1990 to 2011 places TWA 30A well below typical X-ray luminosities for low-mass, pre-MS M type stars.  An important difference to consider when comparing TWA 30A with stars of similar spectral type in the ONC and IC 348 is stellar age and evolution.  While TWA 30A may still host a circumstellar disk with signatures of youth, its age ($\sim$ 8 Myr) is markedly older than its counterparts in the ONC and IC 348 \citep[mean age of 1-2 and 2-3 Myr, respectively][]{Hillenbrand1997,Stelzer2012}.  Moreover, with a spectral type of  M5/M6 (see Section 3) and age of $\sim$8 Myr, TWA 30A lies very close to the substellar boundary as indicated by its HR diagram position given pre-MS stellar evolution models of \citet{Baraffe1998} and \citet{Dantona1997} (e.g., see Figures 5 and 1 in \citealt{Luhman2003} and \citealt{Preibisch2005b} for HR diagrams and evolutionary models of pre-MS stars in Taurus and pre-MS brown dwarfs in the ONC, respectively).  Such models indicate an M5/M6 at $\sim$8 Myr may have evolved from and will evolve to a later spectral type ($\sim$M7, i.e., the substellar boundary).  Indeed, while the X-ray luminosity of TWA 30A is lower than its $\sim$M5 counterparts in IC 348 \citep{Stelzer2012}, it is consistent with those of the $\sim$ 1 Myr ONC $\sim$M7 brown dwarfs \citep{Preibisch2005b}.

Fractional X-ray luminosity is a commonly used proxy for magnetic activity in young stars [e.g.,][]\citep{Grosso2007}.  TWA 30A's fractional X-ray luminosity in the 2011 July epoch ($L_{X}$/$L_{bol}$ $\sim 1.0\times10^{-4}$) was lower than many of the reported $\sim$ M5-M7 pre-MS stars in the ONC, Taurus, and IC 348 \citep{Grosso2007,Gudel2007,Stelzer2012} while its intrinsic $L_{X}$ is significantly fainter than its similar-massed counterparts.  If these values are representative of TWA 30A's quiescent state, the low $L_{X}$/$L_{bol}$ and $L_{X}$ values suggest that TWA 30A may be a representative of the progenitors of the late-type M and early L type {\it main-sequence} dwarf stars which display a lack of X-rays \citep{Berger2006,Berger2010}.  In the case of main sequence stars, such phenomena may be the result of changes in magnetic field configuration or the decreasing ionization fractions in the stellar atmosphere \citep{Berger2006, Mohanty2002}. A smaller ionization fraction in the stellar atmosphere will decrease the coupling of that material to the stellar magnetic fields and, thus, will decrease the coronal X-ray luminosity. With a spectral type of M5/M6, TWA 30A \citep[T$_{eff}$ $\sim$3050;][]{Luhman2003} lies at the border of the regime (T$_{eff}$ of 1500-3000)  investigated in \citet{Mohanty2002} for main sequence mid M to late L stars, where magnetic interaction (between the chromosphere and stellar magnetic field) was limited.

It also follows that TWA 30B, with an upper limit of $L_{X}<3.0\times10^{27}$ erg s$^{-1}$ appears unusually X-ray faint compared to similar stars in the ONC, Taurus and IC 348.  Specifically, given a spectral type of $\sim$M4 and an age of $\sim$ 8 Myr, we estimate the fractional X-ray luminosity of TWA 30B of $L_{X}$/$L_{bol}$ $< 1.5\times10^{-5}$ \citep{Luhman2003}, a value well below that of its similar mass counterparts in other nearby star-forming regions.  Additional X-ray observations of pre-MS mid-M stars (e.g., pre-MS stars that will evolve into main sequence M7 and early L dwarfs) are needed to better understand whether such a decrease in magnetic activity as probed by $L_{X}/L_{bol}$ is characteristic of these young low-mass stars near the H burning limit.  TWA 30A and 30B appear to be at the cusp of this X-ray active/inactive boundary.  This suggests that, in this binary, we could be seeing the \mydoubleq{turn-off} of X-ray activity in pre-MS low mass stars that is observed in main sequence stars of $\sim$M7 or later.

While optical evidence suggests TWA 30A is actively accreting \citep{Looper2010a}, we find no evidence of shock-heated plasma at temperatures characteristic of free-fall accretion onto a mid M-type pre-MS star \citep[T$\sim$2 MK;][and ref. therein]{Brickhouse2010}.  This suggests the hydrogen absorbing column density ($N_{H}$) is too large and/or the emission measure of shocked material is too small to detect such an X-ray accretion signature.  To probe shocks at the base of a jet, i.e., shocks farther from the star and thus more difficult to hide behind an absorbing disk, higher resolution X-ray observations are needed.

\subsection{Relationship Between Atomic and Optical Extinction}

A comparison of measurements of $N_{H}$ and $A_{V}$ for TWA 30A at our two observing epochs (see Tables 2 and 3) is shown in Figure \ref{twa30_nh_av}.  These results are overlaid with $N_{H}$ vs. $A_{V}$ curves determined for the ISM \citep{Ryter1996} and $\rho$ Ophiuchi molecular cloud gas \citep{Vuong2003}. Regardless of the choice of $R_{V}$ assumed in our calculation of $A_{V}$, a correlation between optical and X-ray extinction is evident for TWA 30A: both $A_{V}$ and $N_{H}$ are found to decrease between 2011 June to 2011 July.  Such variability is consistent with variable obscuration of the photosphere by a disk warp and/or clump.  Other pre-MS star-disk systems notable for their high disk inclinations show similar levels of optical variability \citep[e.g., AA Tau; RY Lupi; T Cha;][]{Grosso2007,Manset2009,Schisano2009} although no correlated X-ray emission has yet been detected with contemporaneous observations.

In the case of TWA 30A, an estimate of the warp/clump mass can be made from the X-ray spectral fitting results if we assume that this material completely covers the photosphere of the star during the 2011 June observation and does not cover the photosphere during the 2011 July observations.  The clump column density can then be estimated as the difference between the 2011 June and July observations ($7.9 \times 10^{21}$ cm$^{-2}$).  The clump mass is then $M_{cl}$ = $N_{H}$$\times \pi \times$ $R_{*}^{2}$ $\times$ $m_{H}$.  Using the effective temperature, bolometric luminosity of TWA 30A ($\sim$3050 K and $7.6 \times 10^{31}$ erg s$^{-1}$) and the Stefan-Boltzmann law, we estimate the radius of TWA 30A to be $R_{*}$ $\sim$0.5 R$_{\odot}$.  We hence infer a clump mass of M$_{cl}$ $\approx$ $5.0 \times 10^{19}$ g or $8.4 \times 10^{-9} M_{\oplus}$.  A similar estimate was made for a disk clump in T Cha, for which \citet{Schisano2009} report a clump mass $M_{cl}$ $\approx$ $4 \times 10^{20} g$. However, their calculation was based on optical extinction ($A_{V}$) and an assumed $N_{H}$ to $A_{V}$ relationship.   For comparison, the mass of the Martian moon Phobos is $1.06 \times 10^{19}$ $g$ \citep{Patzold2014}. 

Independent measurements of $N_{H}$ and $A_{V}$  can be used to determine the ratio of atomic absorption from gas and optical extinction due to dust \citep[e.g., see][]{Sacco2014}.  Comparing these values as determined for the nearly edge-on circumstellar disk of TWA 30A with those typical of the ISM (Figure \ref{twa30_nh_av}) can help constrain the chemical environment in which planets might form.  In this respect, it is intriguing that the apparent correlated changes in $N_{H}$ and $A_{V}$ we find for the TWA 30A disk follow the trend of increasing $N_{H}$ with increasing $A_{V}$ displayed in Figure \ref{twa30_nh_av} for the ISM and stars in $\rho$ Oph. However, even considering different values of $R_{V}$, at both epochs TWA 30A lies below the line typical of that of the ISM, indicating a deficiency of the metals (C, N, and/or O) that are most responsible for soft X-ray absorption.  This suggests that either these metals are not abundant in the gas component of the TWA 30A disk, or that these metals are frozen out as molecular (e.g., CO, CO$_{2}$, CH$_{4}$) ices onto dust grains, reducing their X-ray absorption cross-sections.  Such molecular gas freeze-out has long been inferred in T Tauri Star circumstellar disks \citep[e.g., ][]{Dutrey1997} at radii larger than the  $''$snow line$''$ (freeze-out radius) for each chemical species. Recent observations have confirmed the presence of such a snow line, for CO, within the disk orbiting  TW Hya \citep{Qi2013}.  However, we caution that the $N_{H}$/$A_{V}$ values typical of that of the ISM depend on dust grain properties that are not necessarily characteristic of the TWA 30A disk.  Furthermore, standard X-ray absorbing models assume conditions similar to that of the ISM, where only $\sim20\%$ of available gas is in molecular form \citep{Wilms2000}.  In circumstellar disks, the percentage of gas in molecular form is likely to be far higher, and as noted, the gas phase metals may be depleted.

\section{Summary}

We have used contemporaneous multiwavelength observations to study two of the nearest known examples of actively accreting, pre-MS star systems, TWA 30A and 30B.  These two stars have masses near the substellar boundary and are orbited by circumstellar disks viewed nearly edge-on, with evidence for collimated stellar outflows.  Both TWA 30A and 30B have previously displayed highly variable near-infrared variability on timescales of hours to days.  We obtained and analyzed two epochs of XMM-Newton X-ray observations contemporaneous with VLT XSHOOTER and IRTF SpeX spectra to investigate variability, magnetic activity, and the composition of circumstellar disk material associated with TWA 30A and 30B.

We measure TWA 30A at both high and low levels of extinction ($A_{V}$= 14.9, and 4.7).  While the lower level of extinction is consistent with previous measurements, the highest extinction we measure is several magnitudes higher than previously reported. We also report a previously unobserved near-IR excess during both 2011 June and July epochs.  For TWA30B, we measure variable near-IR excesses that are consistent with the low end of the range measured in previous studies. From X-ray spectral fitting of TWA 30A, we find intrinsic 0.15-10.0 keV X-ray luminosities of $\sim$  $2 \times10^{28}$ and $\sim$ $8 \times10^{27}$ erg s$^{-1}$ in 2011 June and July, respectively, and column densities of $N_{H}$ $\sim$ $1 \times10^{22}$ and $N_{H}$ $\sim$ $2 \times10^{21}$ cm$^{-2}$, respectively.  The 0.15-2.0 keV X-ray luminosities of TWA 30A display a factor of 20-100 decrease relative to a 1990 ROSAT X-ray detection. TWA 30B was not detected by XMM-Newton and an upper limit of $L_{X}$ $<$ 3.0 $\times$10$^{27}$ erg s$^{-1}$  was estimated. Magnetic activity, as probed by $log(L_{X}$/$L_{bol}$), is measured for TWA 30A from the 2011 July epoch ($\sim$ $-4.0$) and estimated for TWA 30B ($\leq$ $-4.8$), and are relatively lower than many of the younger pre-MS stars of similar mass in the ONC, Taurus and IC 348. Combined with their very faint intrinsic X-ray luminosities, this suggests that, in the TWA 30 binary, we could be seeing the onset of the decline in ultra-low-mass star X-ray activity that is observed in main sequence stars of $\sim$ M7 or later.  No soft X-ray excess indicative of emission due to shocks, resulting from accretion or jet activity, was detected.

Near-IR spectra were analyzed to estimate disk dust extinction and, combined with XMM-Newton spectra, to measure the ratio of $N_{H}/A_{V}$ in the circumstellar disk of TWA 30A.  We find that variations in optical/IR extinction due to dust and X-ray absorption due to gas appear to be correlated.  We estimate the mass obscuring the photosphere during the 2011 June observation to be $M_{cl} \approx 5.0 \times 10^{19} g$.  Assuming an $R_{V}=3.1$, we measure $N_{H}$ to $A_{V}$ ratios of $\sim$3 $\times$ $10^{20}$ and $\sim$6 $\times$ $10^{20}$ mag$^{-1}$cm$^{-2}$ for observations during low and high extinction states, respectively. These values are lower than those typical of the ISM, suggesting the circumstellar disk metals responsible for absorbing X-rays are either not present or are frozen out onto dust grains. 
 
\section*{Acknowledgments}
We thank D. Looper for providing IRTF observations of TWA 30, the ESO director for providing DDT XSHOOTER observations of TWA 30A and 30B, and R. Montez Jr. and V. Rapson for helpful suggestions during the data reduction process.  This research was supported in part by grants to RIT from the NASA GSFC XMM-Newton Guest Observer Facility (award NNX12AB63G), the National Science Foundation (award AST--1108950), and the NASA Astrophysics Data Analysis Program (award NNX12AH37G).  D. Principe acknowledges a CONICYT-FONDECYT award (grant 3150550); and the Millennium Science Initiative (Chilean Ministry of Economy; grant Nucleus RC 130007).




\bibliographystyle{mnras}






\bsp	
\label{lastpage}
\end{document}